\documentclass[12pt]{article}

\catcode`\@=11

\@addtoreset{equation}{section}
\def\theequation{\thesection\arabic{equation}}

\def\@normalsize{\@setsize\normalsize{15pt}\xiipt\@xiipt
\abovedisplayskip 14pt plus3pt minus3pt%
\belowdisplayskip \abovedisplayskip
\abovedisplayshortskip  \z@ plus3pt%
\belowdisplayshortskip  7pt plus3.5pt minus0pt}
\def\small{\@setsize\small{13.6pt}\xipt\@xipt
\abovedisplayskip 13pt plus3pt minus3pt%
\belowdisplayskip \abovedisplayskip
\abovedisplayshortskip  \z@ plus3pt%
\belowdisplayshortskip  7pt plus3.5pt minus0pt
\def\@listi{\parsep 4.5pt plus 2pt minus 1pt
            \itemsep \parsep
            \topsep 9pt plus 3pt minus 3pt}}

\def\underline#1{\relax\ifmmode\@@underline#1\else
        $\@@underline{\hbox{#1}}$\relax\fi}
\@twosidetrue \relax

\catcode`@=12

\catcode`\@=11

\def\section{\@startsection{section}{1}{\z@}{3.5ex plus 1ex minus
   .2ex}{2.3ex plus .2ex}{\large\bf}}
\def\thesection{\arabic{section}.}

\def\ps@headings{\def\@oddfoot{}\def\@evenfoot{}
\def\@oddhead{\hbox{}\hfill
        \makebox[.5\textwidth]{\raggedright\ignorespaces --\thepage{}--
        \hfill }}
\def\@evenhead{\@oddhead}
\def\subsectionmark##1{\markboth{##1}{}}
}

\ps@headings

\catcode`\@=12

\relax

\def\figcap{\section*{Figure Captions\markboth
        {FIGURECAPTIONS}{FIGURECAPTIONS}}\list
        {Fig. \arabic{enumi}:\hfill}{\settowidth\labelwidth{Fig. 999:}
        \leftmargin\labelwidth
        \advance\leftmargin\labelsep\usecounter{enumi}}}
 \relax
\def\tablecap{\section*{Table Captions\markboth
        {TABLECAPTIONS}{TABLECAPTIONS}}\list
        {Table \arabic{enumi}:\hfill}{\settowidth\labelwidth{Table 999:}
        \leftmargin\labelwidth
        \advance\leftmargin\labelsep\usecounter{enumi}}}
 \relax
\def\reflist{\section*{References\markboth
        {REFLIST}{REFLIST}}\list
        {[\arabic{enumi}]\hfill}{\settowidth\labelwidth{[999]}
        \leftmargin\labelwidth
        \advance\leftmargin\labelsep\usecounter{enumi}}}
 \relax

\catcode`\@=11

\def\marginnote#1{}
\newcount\hour
\newcount\minute
\newtoks\amorpm
\hour=\time\divide\hour by60 \minute=\time{\multiply\hour by60
\global\advance\minute by- \hour}
\edef\standardtime{{\ifnum\hour<12 \global\amorpm={am}%
    \else\global\amorpm={pm}\advance\hour by-12 \fi
    \ifnum\hour=0 \hour=12 \fi
    \number\hour:\ifnum\minute<100\fi\number\minute\the\amorpm}}
\edef\militarytime{\number\hour:\ifnum\minute<100\fi\number\minute}
\def\draftlabel#1{{\@bsphack\if@filesw {\let\thepage\relax
  \xdef\@gtempa{\write\@auxout{\string
    \newlabel{#1}{{\@currentlabel}{\thepage}}}}}\@gtempa
    \if@nobreak \ifvmode\nobreak\fi\fi\fi\@esphack}
     \gdef\@eqnlabel{#1}}
\def\@eqnlabel{}
\def\@vacuum{}
\def\draftmarginnote#1{\marginpar{\raggedright\scriptsize\tt#1}}
\def\draft{\oddsidemargin -.5truein
        \def\@oddfoot{\sl preliminary draft \hfil
        \rm\thepage\hfil\sl\today\quad\militarytime}
        \let\@evenfoot\@oddfoot \overfullrule 3pt
        \let\label=\draftlabel
        \let\marginnote=\draftmarginnote

\def\@eqnnum{(\theequation)\rlap{\kern\marginparsep\tt\@eqnlabel}%
\global\let\@eqnlabel\@vacuum}  }
\def\preprint{\twocolumn\sloppy\flushbottom\parindent 1em
        \leftmargini 2em\leftmarginv .5em\leftmarginvi .5em
        \oddsidemargin -.5in    \evensidemargin -.5in
        \columnsep 15mm \footheight 0pt
        \textwidth 250mmin      \topmargin  -.4in
        \headheight 12pt \topskip .4in
        \textheight 175mm
        \footskip 0pt

\def\@oddhead{\thepage\hfil\addtocounter{page}{1}\thepage}
        \let\@evenhead\@oddhead \def\@oddfoot{} \def\@evenfoot{}
}
\def\titlepage{\@restonecolfalse\if@twocolumn\@restonecoltrue\onecolumn
     \else \newpage \fi \thispagestyle{empty}\c@page\z@
        \def\thefootnote{\fnsymbol{footnote}} }
\def\endtitlepage{\if@restonecol\twocolumn \else  \fi
        \def\thefootnote{\arabic{footnote}}
        \setcounter{footnote}{0}}  
\catcode`@=12 \relax

\def\ps@headings{\def\@oddfoot{}\def\@evenfoot{}
\def\@oddhead{\hbox{}\hfill
        \makebox[.5\textwidth]{\raggedright\ignorespaces --\thepage{}--
        \hfill }}
\def\@evenhead{\@oddhead}
\def\subsectionmark##1{\markboth{##1}{}}
}

\ps@headings

\relax

\usepackage{graphicx}

\begin{document}

\begin{titlepage}
\begin{flushright}

NTUA--7--01 \\
 hep-th/0111052

\end{flushright}

\begin{centering}
\vspace{.41in}
{\large {\bf Cosmological Evolution of a Brane Universe in a Type
0 String Background.}}\\

\vspace{.2in}

 {\bf I. Pappa$^{a} $}  \\
\vspace{.2in}

 National Technical University of Athens, Physics
Department, Zografou Campus,\\ GR 157 80, Athens, Greece. \\

\vspace{0.5in}

{\bf Abstract} \\

\end{centering}

\vspace{.1in}
We study the cosmological evolution of a D3-brane Universe in a
type 0 string background. We follow the brane universe along the
radial coordinate of the background and we calculate the energy
density which is induced on the brane because of its motion in the
bulk. For constant values of tachyon and dilaton an inflationary
phase is appearing. For non constant values of tachyon and dilaton
and for a particular range of values of the scale factor of the
brane-universe, the effective energy density is dominated by a
term proportional to $\frac{1}{(\log\alpha)^{4}}$ indicating a
slowly varying inflationary phase.

\vspace{0.5in}
\begin{flushleft}

   $^{a} $ e-mail
address:gpappa@central.ntua.gr

\end{flushleft}

\end{titlepage}

\newpage
\section{Introduction}

 There has been much recent interest in the idea that our universe
 may be a brane embedded in some higher dimensional space
 \cite{Reg}. It has been shown that the hierarchy problem can be
 solved if the higher dimensional Planck scale is low and the
 extra dimensions large \cite{Dim}. Randall and Sundrum \cite{Rand}
 proposed a solution of the hierarchy problem
 without the need for large extra dimensions but instead through
 curved five-dimensional spacetime $AdS_{5}$ that generates an
 exponential suppression of scales.

 This idea of a brane-universe can naturally be applied to string
 theory. In this context, the Standard Model gauge bosons as well as
 charged matter arise as fluctuations of the D-branes. The universe
 is living on a collection of coincident branes, while gravity and
 other universal interactions is living in the bulk space
 \cite{Pol}.

This new concept of brane-universe naturally leads to a new
 approach to cosmology. Any cosmological evolution like inflation
 has to take place on the brane while gravity acts globally on
 the whole space.

 Another approach to cosmological evolution of our brane-universe
 is to consider the motion of the brane in  higher
 dimensional spacetimes. In \cite{Cha} the motion of a domain wall
 (brane) in such a space  was studied. The Israel matching conditions
 were used to relate the bulk to the domain wall (brane) metric, and
 some interesting cosmological solutions were found. In
 \cite{Keh} a universe three-brane is considered in motion
 in ten-dimensional space in the presence of a gravitational field
 of other branes. It was shown that this motion in ambient space
 induces cosmological expansion (or contraction) on our universe,
 simulating various kinds of matter.

 In \cite{Papa} we examine the motion of a three-brane in a
 background of the type 0 string theory.
 Employing the technics of ref. \cite{Keh}, we will show that a cosmological evolution is induced
 on the three-brane as it moves in the type 0 string background,
 which for some range of the parameters has an inflationary epoch. As we will
 discuss in the following, the tachyon function $f(T)$ which couples to
 the $F_{5}$ form of the type 0 strings, is crucial for the
 inflationary evolution of the brane-universe.

 Type 0 string theories \cite{Tset}
 are interesting because of their connection
 \cite{Typ0} to four-dimensional
 $SU(N)$ gauge theory. The type 0 string does not have spacetime
 supersymmetry and because of that contains in its spectrum a
 non-vanishing tachyon field. In \cite{Tset} it was argued that one
 could construct the dual of an SU(N) gauge theory with 6 real
 adjoint scalars by stacking N electric D3 branes
 of the type 0 model on top of each other. The tachyon field
 couples to the five form field strength, which drives the tachyon
 to a nonzero expectation value.

 Asymptotic solutions of the dual gravity background were
 constructed in \cite{Tset,Minah}. At large radial coordinate
 the tachyon is constant and one
 finds a metric of the form $AdS_{5} \times S^{5}$ with vanishing
 coupling which was interpreted as a UV fixed point. The solution
 exhibits a logarithmic running in qualitative agreement with the
 asymptotic freedom property of the field theory. At small radial
 coordinate the tachyon vanishes and one finds again a solution of the
 form $AdS_{5} \times S^{5}$ with infinite coupling, which was
 interpreted as a strong coupling IR fixed point.
 A gravity solution which describes the flow from the UV
 fixed point to the IR fixed point is given in \cite{Magg}.

 In \cite{Pappa} we study the cosmological evolution of the brane-universe as the brane
 moves from the UV to the IR fixed point. In the following we will calculate
 the effective energy density which is induced
 on the brane because of its motion in the particular background
 of a type 0 string. Using the approximate solutions of
 {\cite{Tset,Minah,Magg}, we find that for large values of the
 radial coordinate $r$, in the UV region, the effective energy density
 takes a constant value, which means that the universe has an
 inflationary period. For smaller values of $r$, or of the scale factor
 $\alpha$, the energy  density is dominated by a term proportional to
 $ \frac{1}{(\log \alpha)^{4}} $, where $\alpha$ is the scale factor of the
 brane-universe. This value of the energy density indicates that
 the universe is in a slow inflationary phase,
 in a "logarithmic inflationary" phase as we can call it, in contrast to
 "constant inflationary" phase which characterizes the usual
 exponential behaviour.
 For even smaller values of $r$,
 the  approximation breaks down and we cannot trust
 the solutions anymore. If we go to the IR region the energy density is
 dominated by the term  $ \frac{1}{\alpha^{4}} $ and again
 we find the "logarithmic inflation" for larger values of $r$. The
 approximation breaks down again for some larger values of $r$. It is well
 known that it is very difficult to connect the IR to the UV
 solutions. Therefore our failure to present a full cosmological
 evolution, relies exactly on this fact .

 We note here that what we find is somewhat peculiar, in the sense that one
 does not expect the effective energy density to be dominated, for
 a range of values of the scale factor, by terms proportional to
 $ \frac{1}{(\log \alpha)^{4}}$.  We understand this behaviour, as due entirely
 to mirage matter which is induced on the brane, from this particular background.

 Our work is organized as follows. In section two, we briefly
 review the technics of ref. \cite{Keh}. In section three, using
 this solution we find the cosmological evolution of the
 three-brane in a background of type 0 string with constant dilaton and tachyon. In section four
 we study the cosmological evolution of the
 three-brane in a background of type 0 string with non constant dilaton
 and tachyon. Finally in the last
 section  we discuss our results.

\section{Brane Universe}

We will consider a probe brane moving in a generic static,
 spherically symmetric background \cite{Keh}. We assume the brane
 to be light compared to the background so that we will neglect
 the back-reaction. As the brane moves the induced world-volume
 metric becomes a function of time, so there is a cosmological
 evolution from the brane point of view. The metric of a D3-brane
 is parameterized as
\begin{equation}\label{in.met}
ds^{2}_{10}=g_{00}(r)dt^{2}+g(r)(d\vec{x})^{2}+
  g_{rr}(r)dr^{2}+g_{S}(r)d\Omega_{5}
\end{equation}
 and there is also a dilaton field $\Phi$ as well as a $RR$
 background $C(r)=C_{0...3}(r)$ with a self-dual field strength. The
 action on the brane is given by
\begin{eqnarray}\label{B.I. action}
  S&=&T_{3}~\int~d^{4}\xi
  e^{-\Phi}\sqrt{-det(\hat{G}_{\alpha\beta}+(2\pi\alpha')F_{\alpha\beta}-
  B_{\alpha\beta})}
   \nonumber \\&&
  +T_{3}~\int~d^{4}\xi\hat{C}_{4}+anomaly~terms
\end{eqnarray}
 At last we get for the induced metric on the brane
\begin{equation}\label{fin.ind.metric}
d\hat{s}^{2}=-d\eta^{2}+g(r(\eta))(d\vec{x})^{2}
\end{equation}
 with $\eta$ the cosmic time.

This equation is the standard form of a flat expanding universe.
If we define the scale factor as $\alpha^{2}=g$ then we can
interpret the quantity $(\frac{\dot{\alpha}}{\alpha})^{2}$ as an
effective matter density on the brane with the result
\begin{equation}\label{dens}
\frac{8\pi}{3}\rho_{eff}=\frac{(C+E)^{2}g_{S}e^{2\Phi}-|g_{00}|(g_{S}g^{3}+\ell^{2}e^{2\Phi})}
{4|g_{00}|g_{rr}g_{S}g^{3}}(\frac{g'}{g})^{2}
\end{equation}

Therefore the motion of a D3-brane on a general spherically
symmetric background had induced on the brane a matter density. As
it is obvious from the above relation, the specific form of the
background will determine the cosmological evolution on the brane.

\section{Brane-inflation}
We consider a D3-brane moving along a geodesic in the background
of a type 0 string. For the case of constant tachyon and dilaton
the solution for the induced metric on the brane is \cite{Tset}
\begin{equation}\label{Sol}
  g_{00}=-H^{-\frac{1}{2}},
  g(r)=H^{-\frac{1}{2}},  g_{S}(r)=H^{\frac{1}{2}}r^{2},
g_{rr}(r)=H^{\frac{1}{2}},    H=1+\frac{e^{\Phi_{0}}Q}{2r^{4}}
\end{equation}
the four form is
\begin{equation}\label{Cterm}
C=e^{-\Phi_{0}}f^{-1}(T)(1+\frac{e^{\Phi_{0}}Q}{2r^{4}})^{-1}+Q_{1}
\end{equation}
where $Q_{1}$ is a constant and f(T) is the function
\begin{equation}\label{ftac}
 f(T)=1+T+\frac{1}{2} T^{2}
\end{equation}
by which the tachyon is coupled to the $RR$ field.

 The effective density on the brane
(\ref{dens}), using eq.(\ref{Sol}) and (\ref{Cterm})  becomes
\begin{equation}\label{cre}
\frac{8\pi}{3}\rho_{eff}=\frac{1}{4}[(f^{-1}(T)+EHe^{\Phi_{0}})^{2}-(1+\frac{\ell^{2}e^{
2\Phi_{0}}}{2}H)]
\frac{Q^{2}e^{2\Phi_{0}}}{r^{10}}H^{-\frac{5}{2}}
\end{equation}
where the constant $Q_{1}$ was absorbed in a redefinition of the
parameter $E$. Identifying $g=\alpha^{2}$ and using
$g=H^{-\frac{1}{2}}$ we get from (\ref{cre})
\begin{eqnarray}\label{aro}
\frac{8\pi}{3}\rho_{eff}&=&(\frac{2e^{-\Phi_{0}}}{Q})^{\frac{1}{2}}
 \Big{[} \Big{(} f^{-1}(T)+\frac{Ee^{\Phi_{0}}}{\alpha^{4}} \Big{)}^{2}
-\Big{(}1+\frac{\ell^{2}e^{2\Phi_{0}}}
{\alpha^{6}}(\frac{2e^{-\Phi_{0}}}{Q})^{\frac{1}{2}}\nonumber \\&&
(1-\alpha^{4})^{\frac{1}{2}} \Big{)}  \Big{]} (1-\alpha^{4})
^{\frac{5}{2}}
\end{eqnarray}
From the relation $g=H^{-\frac{1}{2}}$ we find
\begin{equation}\label{ro}
  r= (\frac{\alpha^{4}}{1-\alpha^{4}})
  ^{\frac{1}{4}}(\frac{Qe^{\Phi_{0}}}{2})^{\frac{1}{4}}
\end{equation}
This relation restricts the range of $\alpha$ to $0\leq \alpha
<1$, while the range of $r$ is $0\leq r< \infty$. We can calculate
the scalar curvature of the four-dimensional universe as
\begin{equation}\label{curv}
  R_{brane}=8\pi(4+\alpha\partial_{\alpha})\rho_{eff}
\end{equation}
If we use the effective density of (\ref{aro}) it is easy to see
that $R_{brane}$ of (\ref{curv}) blows up at $\alpha=0$. On the
contrary if $r\rightarrow 0$,then the $ds^{2}_{10}$ of
(\ref{in.met}) becomes
\begin{equation}\label{ads}
ds^{2}_{10}= \frac{r^{2}}{L} (-dt^{2}+(d\vec{x})^{2})+
      \frac{L}{r^{2}} dr^{2}+  L d\Omega_{5}
\end{equation}
with $L=(\frac{e^{\Phi_{0}}Q}{2})^{\frac{1}{2}}$. This space  is a
regular $AdS \times S^{5}$ space.

Therefore the brane develops an initial singularity as it reaches
$r=0$, which is a coordinate singularity and otherwise a regular
point of the $AdS_{5}$ space. This is another example in Mirage
Cosmology \cite{Keh} where we can understand the initial
singularity as the point where the description of our theory
breaks down.

If we take $\ell^{2}=0$, set the function $f(T)$ to each minimum
value and also taking $\Phi_{0}=0$, the effective density
(\ref{aro}) becomes
\begin{equation}\label{laro}
\frac{8\pi}{3}\rho_{eff}=(\frac{2}{Q})^{\frac{1}{2}}
\Big{(}(2+\frac{E}{\alpha^{4}})^{2} -1 \Big{)} (1-\alpha^{4})
^{\frac{5}{2}}
\end{equation}
As we can see in the above relation, there is a constant term,
coming from the tachyon function $f(T)$. For small $\alpha$ and
for some range of the parameters $E$ and $Q$ it gives an
inflationary phase to the brane cosmological evolution. In Fig. 1
we have plotted $\rho_{eff}$ as a function of $\alpha$ for $Q=2$.


\begin{figure}[h]
\centering
\includegraphics[scale=0.7]{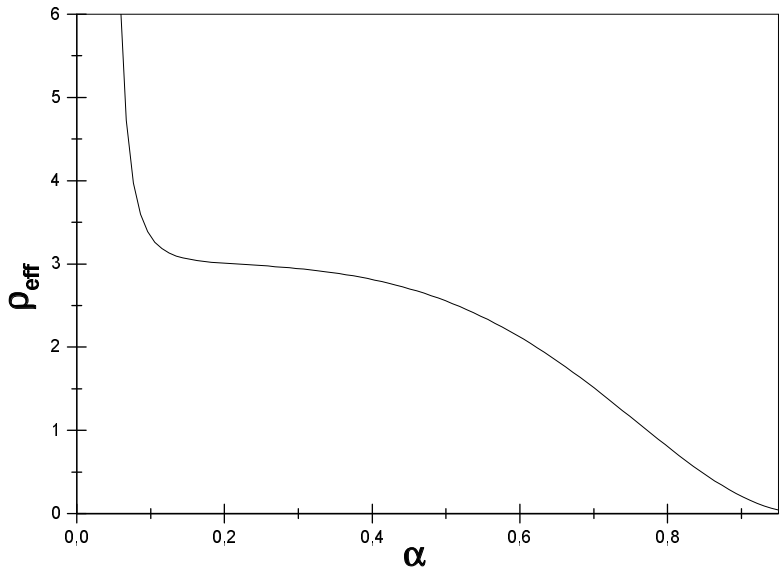}
\caption {The induced energy density on the brane as a function of
the brane scale factor.}
\end{figure}

\section{Cosmological evolution of the Brane-Universe }

In the case of non constant dilaton and tachyon the effective
energy density using the solutions given by Tseytlin, Klebanov and
Minahan \cite{Typ0}, \cite{Minah} is calculated.

We consider again a D3-brane moving along a geodesic in the
background of a type 0 string. Having all the solutions in the
ultra violet and the infrared, we can follow the cosmological
evolution of our universe as it moves along the radial coordinate
$r$. In the presence of a non trivial tachyon field the coupling
$e^{-\Phi}$ which appears in the Dirac-Born-Infeld action in
(\ref{B.I. action}), is modified by a tachyonic function
$\kappa(T)=1+\frac{1}{4}T+O(T^{2})$. Then we can define an
effective coupling \cite{Typ0}
\begin{equation}\label{effphi}
e^{-\Phi}_{eff}=\kappa(T) e^{-\Phi}
\end{equation}

 The bulk
fields are also coordinate dependent and the induced metric on the
brane  will depend on a non trivial way on the dilaton field.
Therefore the metric in the string frame will be connected to the
metric in the Einstein frame through
$g_{St}=e^{\frac{\Phi}{2}}_{eff} g_{E}$.  All the quantities used
so far were defined in the string frame. We will follow our
cosmological evolution in the Einstein frame. Then the relation
(\ref{dens}) becomes
\begin{equation}\label{eindens}
 \frac{8\pi}{3}\rho_{eff}= (\frac {\dot{\alpha}}{\alpha})^{2}=
\frac{(C+E)^{2}g_{S}-|g_{00}|(g_{S}g^{3}+\ell^{2})}
{4|g_{00}|g_{rr}g_{S}g^{3}}(\frac{g'}{g})^{2}
\end{equation}

Having the approximate solution in the UV we can calculate the
metric components and the $RR$ field C. So, we can calculate the
effective energy density from (\ref{eindens}) setting $\ell^{2}=0$
and we get

\begin{eqnarray}\label{rhonimah}
\frac{8\pi}{3}\rho_{eff}& =&
 \Big{[} \Big{(}1-\frac{1}{Q\alpha^{4}}Ei[log2Q + 4log\alpha] +
\frac{E} {2\alpha^{4}} \Big{)}^{2} \nonumber \\ && -
\frac{1}{4}\Big{(}1-\frac{1}{2(log2Q + 4log\alpha)}\Big{)}^{4}
\Big{ ]}\Big{(}1-\frac{1}{2(log2Q +
4log\alpha)}\Big{)}^{-4}\Big{(}1-\frac{9}{2(log2Q +
4log\alpha)}\Big{)}^{-1} \nonumber \\&& \Big{(}1+\frac{1}{(log2Q +
4log\alpha)^{2}}\frac{1}{\Big{(}1-\frac{1}{2(log2Q +
4log\alpha)}\Big{)}}\Big{)}^{2}
\end{eqnarray}

For some typical value of the parameters Q=1 and E=1, and for
large values of $\alpha$, it is obvious that $\rho_{eff}$ has a
constant value. Therefore an observer on the brane will see an
expanding inflating universe. It is interesting to see what
happens for small values of $\alpha$. As $\alpha$ gets smaller, a
term proportional to $\frac{1}{(\log\alpha)^{4}}$ starts to
contribute to $\rho_{eff}$. Therefore the universe for small
values of scale factor has a slow expanding inflationary phase
which we call it "logarithmic inflationary" phase. For smaller
value of $\alpha$ we cannot trust the solution which is reflected
in the fact that $\rho_{eff}$ gets infinite. The behaviour of the
effective energy density as a function of the scale factor is
shown in Figure 2.


\begin{figure}[h]
\centering
\includegraphics[scale=0.9]{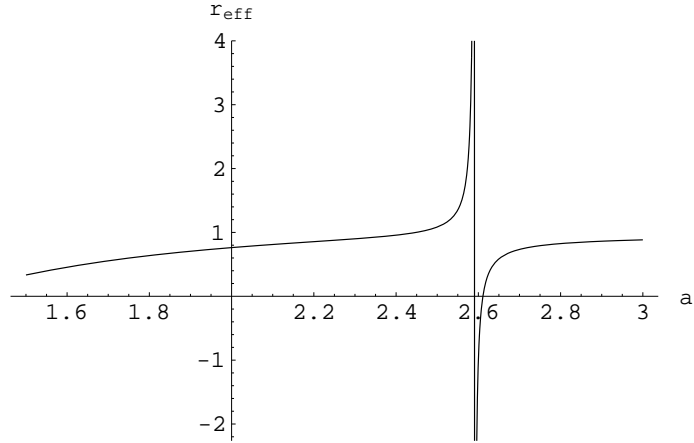}
\caption {The induced energy density on the brane as a function of
the brane scale factor.}
\end{figure}

To have an idea how the slow inflationary phase proceeds, we can
assume for the moment that the effective energy density scales as
\begin{equation}
 \frac{8\pi}{3}\rho_{eff}=(\frac{\dot{\alpha}}{\alpha})^{2}
 =\frac{1}{(\log\alpha)^{p}}
 \end{equation}
The solution of the above equation is
\begin{equation}
\alpha=e^{ \textstyle  t^{ \textstyle (\frac{2}{p+2}) } }
\end{equation}

Therefore we remain in an exponentially growing universe, but
various values of p have the effect of making the universe to slow
down its expansion. We note here that in order to estimate the
behaviour and the duration of this "logarithmic inflationary"
phase, we have to resolve the problem of the singularity.

Going now to IR $\rho_{eff}$ becomes,

\begin{eqnarray}\label{tserho}
\frac{8\pi}{3}\rho_{eff}& =& \Big{[} \Big{
(}-1-2\frac{1}{\sqrt{2}Q\alpha^{4}}Ei[log\sqrt{2}Q + 4log\alpha] +
\frac{E}{\sqrt{2}\alpha^{4}} \Big{)}^{2} - \nonumber
\\ &&
 \frac{1}{2}\Big{(}1+\frac{9}{2(log\sqrt{2}Q +
4log\alpha)}\Big{)}^{4} \Big{]}\Big{(}1+ \frac{9}{2(log\sqrt{2}Q +
4log\alpha)}\Big{)}^{-4} \nonumber \\&&
\Big{(}1+\frac{1}{2(log\sqrt{2}Q + 4log\alpha)}\Big{)}^{-1}
\nonumber
\\&&
\Big{(}1-\frac{9}{2(4log\alpha+ log\sqrt{2}Q)^{2}}
 \frac{1}{\Big{(}1-\frac{9}{2(log\sqrt{2}Q + 4log\alpha)}\Big{)}}\Big{)}^{2}
\end{eqnarray}

As we can see, the above relation is the same as the energy
density in the UV (relation (\ref{rhonimah})) up to some numerical
factors, as expected. The difference is, that now it is valid for
small $\alpha$. For small $\alpha$ first the term $
\frac{1}{\alpha^{8}}$ dominates and then the term
$\frac{1}{\alpha^{4}}$. As $\alpha$ increases the term $
\frac{1}{(\log\alpha)^{4}}$ takes over and drives the universe to
a slow inflationary expansion.

\section{Discussion}

We had followed a probe brane along a geodesic in the background
of type 0 string. Assuming that the universe is described by a
four-dimensional brane, we calculate the effective energy density
which is induced on the brane because of this motion. We study
this mirage matter as the brane-universe moves along the radial
coordinate.

At first we found, that the motion of the brane-universe in this
particular background induces an inflationary phase on the brane.
We made the analysis in the limited case where the dilaton and
tachyon fields were constants. This assumption simplified the
calculation because there is an exact solution of the equations of
motion.

We also extended our study to a background where all the fields
were functions of the radial coordinate. Using the solutions given
by \cite{Typ0}, \cite{Minah}, we calculated the energy densities
that are induced on the brane. What we found is that for large
values of the scale factor as it is measured on the brane (large
values of the radial coordinate) the universe enters a slow
inflationary phase, in which the energy density is proportional to
an inverse power of the logarithm of the scale factor. As the
scale factor grows the induced energy density takes a constant
value and the universe enters a normal exponential expansion. For
small values of the scale factor the induced energy density scales
as the inverse powers of the scale factor and then the logarithmic
terms take over and the universe enters a slow exponential
expansion.

The energy densities we calculated break down for some specific
values of the scale factor. This is a reflection of the fact that
the approximate solutions in the IR cannot be continued to the UV.
To answer the question if there is a true phase of "logarithmic
inflation" in which the universe inflates but with a slow rate, we
must resolve the problem of singularities, where our theory breaks
down.

\end{document}